**Exceptional Optical Phonon Coherence in Enriched Cubic Boron Arsenide via Suppression of Three-Phonon Scattering**

**Authors:** Tong Lin[1,†], Fengjiao Pan[2,†], Gaihua Ye[3], Sanjna Sukumaran[1], Cynthia Nnokwe[3], Ange Benise Niyikiza[2], William A. Smith[1], Stephen B. Bayne[3], Rui He[3,*], Zhifeng Ren[2,*], Hanyu Zhu[1,*]

**Affiliations:**

[1]Department of Materials Science and NanoEngineering, Rice University, Houston, TX, 77005, USA

[2]Department of Physics and Texas Center for Superconductivity at University of Houston (TcSUH), University of Houston, Houston, TX, 77204, USA

[3]Department of Electrical and Computer Engineering, Texas Tech University, Lubbock, TX, 79409, USA

*Corresponding author. Rui He, rui.he@ttu.edu, Zhifeng Ren, zren@uh.edu, Hanyu Zhu, hanyu.zhu@rice.edu

†These authors contributed equally to this work.

**ABSTRACT.** Cubic boron arsenide (BAs) is a promising semiconductor for next-generation electronics due to its outstanding ambipolar mobility and thermal conductivity, the latter of which is attributed to the suppression of three-phonon scattering. However, precisely accounting for different high-order anharmonic scattering processes is challenging from both theory and experiment, so that questions remain open regarding the ultimate limit of phonon lifetime and thermal conductivity in BAs. Here we show that this gap nearly eliminates three-phonon scattering for zone-center optical phonons in a wide temperature range, leading to a record-high, isotope purity-limited phonon coherence with a quality factor above $3.7 \times 10^3$ for >98% enriched $^{11}$BAs below 100 K. We discriminate three decoherence mechanisms by their temperature-dependent contribution to the damping rate using high-resolution Raman and Fourier transform infrared spectroscopy. For the as-synthesized crystals, we find that defect scattering has negligible contributions to the linewidth of optical phonons in comparison to isotope scattering. These results provide critical insights into the intrinsic and extrinsic scattering mechanisms of optical phonons in BAs, motivating further studies to quantify anharmonic effects and realize superior phonon transport.

## I. INTRODUCTION.

The high thermal conductivity of BAs, crucial for high-performance processors and optoelectronics, can be simply understood from the large contrast in atomic mass between B and As. This difference opens a large acoustic-optical gap (a-o gap) and reduces three-phonon scattering [1–8], which is the dominant phonon scattering mechanism in most materials [9]. However, subsequent theoretical studies also suggested that the four-phonon scattering process is not negligible in BAs, leading to a considerable reduction in both intrinsic thermal conductivity and optical phonon lifetime from the initial optimistic estimates [10,11]. Isolating the four-phonon contribution, which is determined by intrinsic phonon anharmonicity, is important for guiding the experimental efforts to mitigate extrinsic phonon scattering and achieve ultrahigh thermal conductivity $\kappa$ [12–15]. To date, most studies have focused on the temperature dependence of thermal conductivity to help differentiate scattering mechanisms primarily for acoustic phonons, where a power law with an exponent larger than 1 suggests a non-negligible four-phonon scattering contribution. First-principles calculations predicted a temperature dependence of $\kappa$ between $1/T^{1.6}$ and $1/T^{1.8}$, arising from a combination of three-phonon, four-phonon, phonon-defect, and phonon-isotope scattering, but experiments showed a broader variation of $1/T^{1.3}$ to $1/T^2$ (300 – 600 K) [1–4,10,11,16]. These discrepancies underscore the theoretical challenges in accurately calculating the four-phonon scattering rates due to computational complexity [13,17] and the strong sensitivity of thermal conductivity to defects [18,19]. The challenges also come from the experimental side regarding the accuracy of extracting thermal conductivity from time-domain thermoreflectance in BAs [15], and anharmonicity of individual phonon modes from such collective measurements of all modes.

Optical phonons, on the other hand, offer a complimentary, and possibly more ideal platform for investigating phonon anharmonicity, because they can be optically excited in a single mode, making it easier to disentangle competing phonon scattering processes [20]. The linewidth of an optical phonon mode in solid crystals directly reflects its total decoherence rate [21], which includes the contribution from temperature-dependent anharmonic scattering, as well as temperature-independent defect scattering and inhomogeneous broadening. Therefore, the optical phonon linewidth as a function of temperature provides direct evidence regarding the strength of various phonon scattering processes in BAs [22]. Furthermore, the combination of its infrared (IR) and Raman activity has enabled multiple experimental approaches for studying optical phonons in BAs. Yet, literature of linewidth measurements in BAs do not have sufficient spectral resolution to accurately resolve the delicate features in BAs, such as the extremely narrow optical phonon linewidth (1 cm$^{-1}$ at room temperature) and small LO and TO splitting (~2 cm$^{-1}$ under ambient condition) [20,22–26], due to large instrumental broadening. A combination of high-resolution phonon spectroscopy and rigorous data analysis is essential for observing the temperature-dependent phonon linewidth and identifying high-order anharmonic effects.

Here we report a comprehensive investigation of the scattering mechanisms of the zone-center optical phonon in isotopically enriched BAs using high-resolution Raman and Fourier transform infrared (FTIR) spectroscopy in ambient and cryogenic temperatures (77-300 K). The unprecedented high resolution enabled the characterization of subtle features like LO-TO splitting and the narrow linewidth. We found that the optical phonon linewidth exhibits a quadratic temperature dependence, indicating that four-phonon scattering is dominant. The results confirm that the large band gap between the acoustic and optical phonons (a-o gap) effectively eliminated three-phonon scattering for optical phonons, leading to an extraordinarily long coherence lifetime at 77 K with a quality factor of $3.7\times10^3$. This contrasts with the case of thermal conductivity, in which the three-acoustic-phonon scattering is still the main limiting factor. The observed temperature-independent residue linewidth of 0.1 cm$^{-1}$ for >98% enriched $^{11}$BAs is compatible with the theoretically expected value from phonon-isotope scattering for such an isotope purity [27]. We also confirm that phonon-defect scattering has a negligible effect on the linewidth of the optical phonons in our samples. Our findings corroborate with earlier first-principles calculations [27], suggesting that for optical phonons in BAs, simply purifying the boron isotope based on current crystal growth conditions has great potential to further extend the coherence lifetime, and provide valuable benchmarks for future theoretical calculations and experimental investigations of phonon anharmonicity in BAs.

## II. METHODS

We synthesized single crystals of BAs using a modified chemical vapor transport (CVT) method as reported previously (Supplemental Material Section 1) [1,4,28,29]. The resulting BAs has a zinc-blende face-centered cubic crystal structure in space group

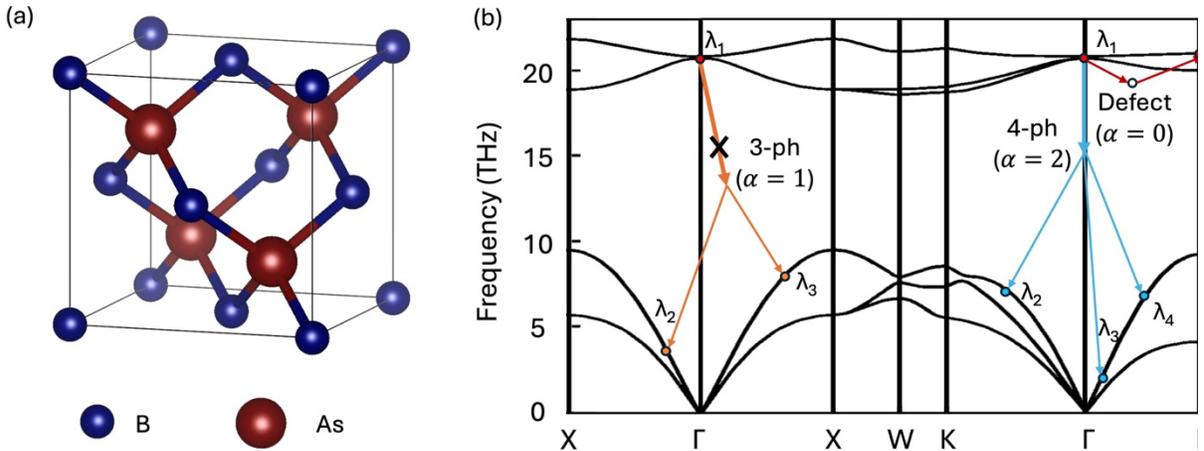

**Figure 1.** Schematics of the crystal structure and phonon scattering mechanisms in BAs. (a) BAs is zinc-blende face-centered cubic in the $F\bar{4}3m$ space group with one formula unit per primitive cell. The heavy atom As mainly affects the acoustic phonon modes while the light atom B controls the optical phonon modes. (b) Phonon band structure of BAs (reproduced from the Materials Project) and the illustration of different types of phonon-scattering mechanisms for zone-center optical phonons in BAs. The three-phonon scattering process ($\gamma \propto T$) is forbidden due to the large energy gap between acoustic and optical phonons and the bunching of acoustic modes. However, the four-phonon scattering process ($\gamma \propto T^2$) and phonon-defect scattering ($\gamma \propto T^0$) are expected to be important.

$F\bar{4}3m$ (Figure 1a). As depicted in Figure 1b, the phonon dispersion in BAs exhibits distinctive features, including the small bandwidth of the optical phonons, the large a-o gap, and the closely bunched acoustic branches [30,31]. Typically, the scattering of optical phonons is predominantly governed by the three-phonon process, in which the scattering rate follows a linear temperature dependence ($\gamma_{3ph} \propto T$), following the population of available low-energy acoustic phonons [32]. However, in isotopically enriched BAs single crystals, the optical phonon at the Γ point cannot readily decay through three-phonon splitting since two other phonon modes that satisfy energy and momentum conservation simultaneously can hardly be found. However, such a constraint is alleviated by the four-phonon scattering process, which has been demonstrated to be remarkably strong in BAs [10,16,22]. Its scattering rate has a quadratic dependence on temperature ($\gamma_{4ph} \propto T^2$), stronger than that for three-phonon scattering since it generally involves two low-energy acoustic phonons. Additionally, BAs has been shown to contain high concentrations of crystal imperfections, including isotopic variations, point defects, and grain boundaries [33,34]. These all result in additional extrinsic scattering channels that are insensitive to temperature [14,35,36]. According to Matthiessen's rule, the overall scattering rate incorporates all relevant mechanisms: $\gamma(T) = \gamma_{3ph}(T) + \gamma_{4ph}(T) + \gamma_{iso} + \gamma_{def}$. Therefore, by analyzing the temperature exponent $\alpha$ of the phonon scattering rate $\gamma$ and the linewidth $\Gamma = \frac{\gamma}{2\pi c} \sim T^\alpha$, valuable information can be obtained regarding the relative contributions of these scattering mechanisms.

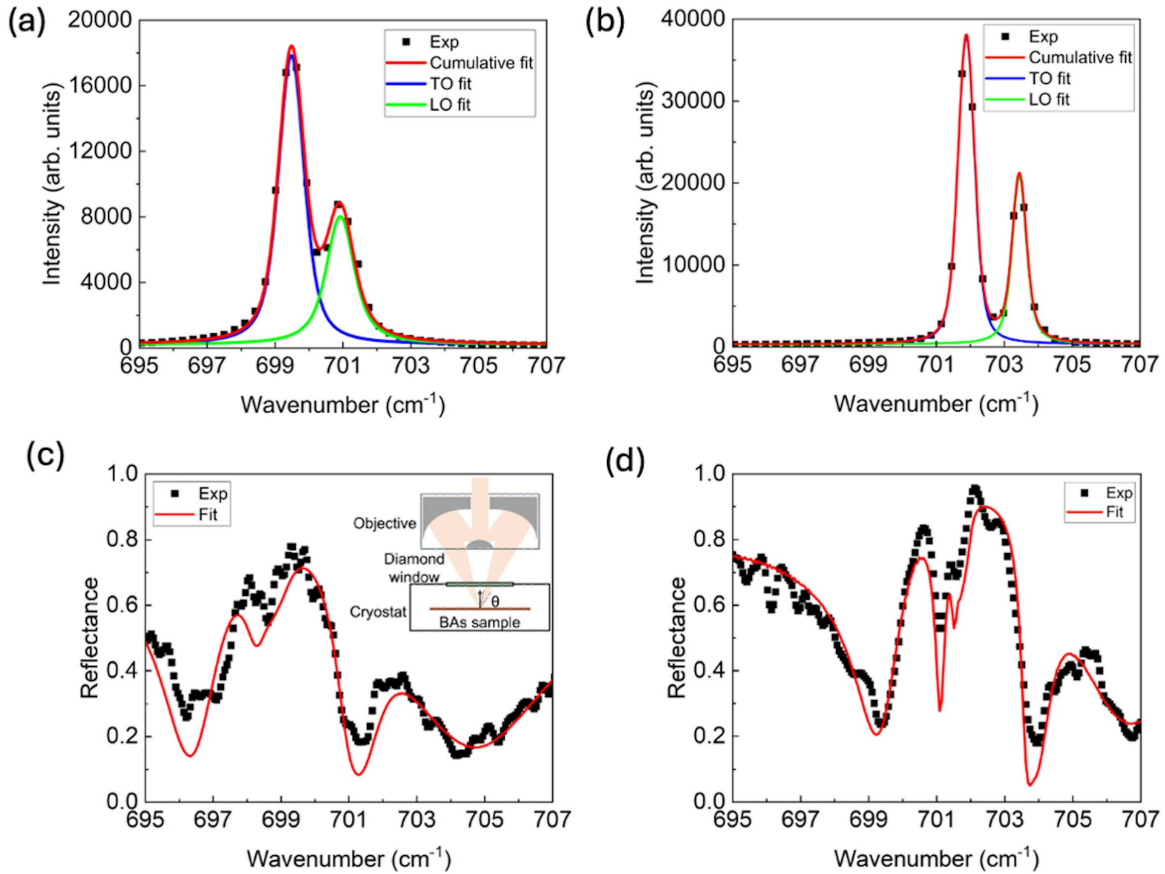

**Figure 2.** High-resolution phonon spectroscopy for BAs at different temperatures. (a, b) Raman spectra for a BAs sample at room temperature (a) and 100 K (b). The experimental data were fitted by Voigt profiles representing TO and LO peaks to extract both frequency and linewidth. The frequencies blueshift and the linewidths narrow with decreasing temperature. (c, d) FTIR reflectance spectra at room temperature (c) and 100 K (d). Phonon frequency and linewidth values can be obtained from the measured reflectance by the harmonic oscillator model of the dielectric function and the transfer matrix method, which agree well with those from Raman spectroscopy. Inset shows the measurement configuration.

## III. RESULTS AND DISCUSSION

Previous investigations of optical phonon linewidth in BAs primarily relied on Raman spectroscopy. However, these efforts were significantly limited by the poor resolution of conventional instruments [20,22–24]. The narrowest linewidth reported was 1.2±0.2 cm[-1] at room temperature for an isotopically enriched [11]BAs sample, a value still much larger than that theoretically predicted [23]. Furthermore, only a single Raman peak, which was assigned as the LO mode, was observed among all measurements under ambient conditions. LO-TO splitting was only observed under high pressure when the splitting exceeded the instrument's resolution [24,37]. In this study, we first utilized a high-resolution Raman setup to characterize the optical phonons in BAs. The line shape of the measured Raman peak is the convolution of the intrinsic Lorentzian phonon line shape with the Gaussian response function of the instrument, which yields a Voigt profile [38]. (Supplemental Material Section 3) [29]. We can obtain the intrinsic Lorentzian full width at half maximum (FWHM) after deconvolution by the instrumental broadening factor using this relation among the FWHM values of the Voigt ($\Gamma_V$), Gaussian ($\Gamma_G$), and Lorentzian ($\Gamma_L$) [39]:

$$\Gamma_V = 0.535\Gamma_L + \sqrt{0.217\Gamma_L^2 + \Gamma_G^2} \qquad (1)$$

Figure 2a and 2b display two representative Raman spectra at room temperature and 100 K, respectively. Benefiting from the enhanced resolution, two strong phonon peaks separated by less than 2 cm[-1], which correspond to the TO and LO modes of the T$_2$ optical phonon [20,22], were observed according to the expected selection rule (Supplemental Material Section 1) [29]. A comparison of the Raman spectra obtained at different temperatures clearly demonstrates phonon blueshift and linewidth narrowing with decreasing temperature. To accurately determine phonon positions and linewidths, two Voigt functions were used to fit the TO and LO peaks. At room temperature, the phonon frequencies and Lorentzian linewidths, i.e., real phonon linewidth $\Gamma_L$ after deconvolution are $\omega_{TO}/2\pi c = 699.43 \pm 0.04$ cm[-1], $\omega_{LO}/2\pi c = 700.91 \pm 0.06$ cm[-1], and $\Gamma_L = 0.66 \pm 0.13$ cm[-1], while at 100 K these values shifted to $\omega_{TO}/2\pi c = 701.68 \pm 0.05$ cm[-1], $\omega_{LO}/2\pi c = 703.32 \pm 0.05$ cm[-1], and $\Gamma_L = 0.20 \pm 0.08$ cm[-1]. This is the first report to resolve the subtle LO-TO splitting and determine intrinsic Lorentzian phonon linewidth. The narrow linewidth of the doubly degenerate TO mode indicates that the phonon population lifetime in arbitrary polarization is at least $1/2\pi c\Gamma_L = 27 \pm 11$ ps, longer than those of the circularly polarized zone-center phonons in 2D materials reported before [21,40].

We further validated our findings using FTIR spectroscopy, which can provide higher spectral resolution than Raman spectroscopy. The mid- to far-IR optical properties originating from optical phonons are important for radiative heat transfer, but have not been reported in single-crystalline BAs so far [41]. We employed FTIR micro-spectroscopy to measure local phonon properties on samples with small areas and modelled the results using transfer matrix method (Supplemental Material Section 4) [29] [4]. However, neither IR microscopy nor the better resolved Raman microscopy found clear spatial variation of the optical phonon properties in our experiments, for reasons discussed later regarding the effect of defects. Two representative reflectance spectra at room temperature and 100 K are shown in Figure 2c and 2d, respectively. The highest reflectance appears between the TO and LO resonances, i.e., the Reststrahlen band, due to the negative refractive index and the large extinction coefficient, and shifted to higher frequencies as the temperature decreased. Additionally, the dip near the TO resonance, which results from dielectric function crossing 1 at a frequency slightly below the TO mode, became sharper with decreasing temperature, suggesting a narrower phonon linewidth (Fig S5). The extracted parameters at room temperature are $\omega_{TO}/2\pi c = 699.25 \pm 0.04$ cm[-1], $\omega_{LO}/2\pi c = 700.95 \pm 0.05$ cm[-1], and $\Gamma = 0.72 \pm 0.09$ cm[-1], while at 100 K, these values are $\omega_{TO}/2\pi c = 702.00 \pm 0.02$ cm[-1], $\omega_{LO}/2\pi c = 703.58 \pm 0.02$ cm[-1], and $\Gamma = 0.19 \pm 0.03$ cm[-1]. Note that the fitted frequencies do not correspond to any of the spectral features, which are caused by the interference of phonon polaritons in the sample plates (Fig S5). These results are consistent with those obtained from Raman measurements, but the linewidth values are more accurate due to better spectral resolution.

Having successfully demonstrated the ability to accurately extract phonon frequencies and linewidths via Raman and FTIR spectroscopy, we subsequently performed detailed temperature-dependent measurements of a BAs sample from room temperature to 100 K at intervals of 25 K using both techniques to identify the scattering mechanisms. Figure 3a shows the nonlinear temperature dependence of the shift in both TO and LO phonon frequencies obtained from both the Raman and FTIR measurements, which agree well with the semi-quantitative model including contributions from thermal expansion and anharmonic scattering (Supplemental Material Section 2) [29].

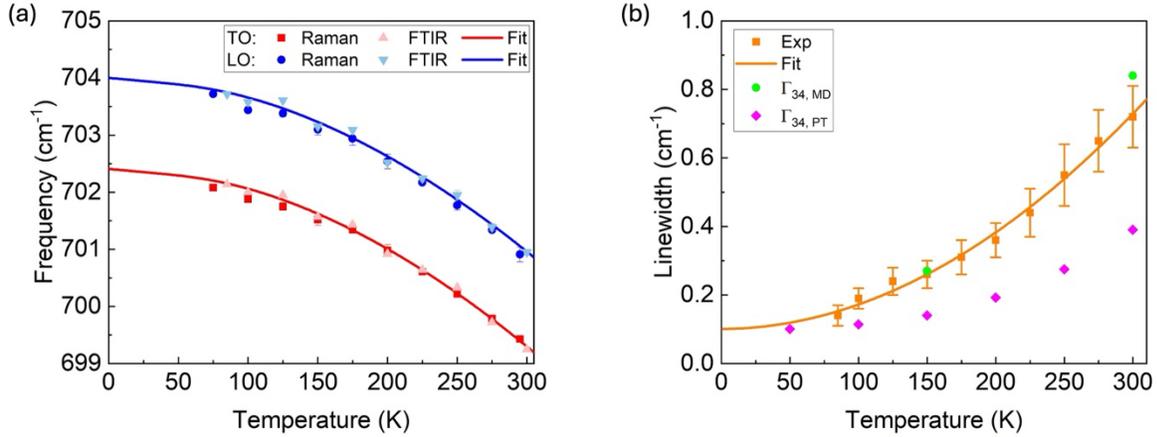

**Figure 3.** Temperature dependence of BAs phonon frequencies and linewidth. (a) Softening of TO and LO phonons at higher temperatures was fitted by a model considering thermal expansion and phonon anharmonicity (including 3-ph and 4-ph scattering) effects. (b) Phonon linewidth as a function of temperature extracted from FTIR measurements and fit with a power-law function ($\gamma = AT^\alpha + B$) that yields an exponent of $\alpha = 2.0$ and a constant of B = 0.10. This quadratic dependence indicates that four-phonon scattering dominates the decay process of optical phonons in BAs. $\Gamma_{34,MD}$ and $\Gamma_{34,PT}$ are obtained from molecular dynamics (ref [13]) and density-functional perturbation theory (ref [27]), respectively, and a constant of 0.1 cm-1 is added to account for the isotope impurity scattering.

Although the model indicates a smaller contribution of 3-phonon scattering than 4-phonon scattering in BAs, more direct evidence is presented by the temperature-dependent linewidth obtained from FTIR (Fig 3b), which have smaller uncertainty compared to those from Raman measurements but are consistent with the latter (Figure S4). The temperature-dependent linewidth values were fit with a power-law function $\Gamma = AT^\alpha + B$ to account for the intrinsic phonon anharmonicity and extrinsic impurity scattering [10]. The fit yielded a result of $\alpha = 2.0 \pm 0.2$ and $B = 0.10 \pm 0.02$ cm$^{-1}$. The temperature exponent $\alpha = 2.0 \pm 0.2$ strongly indicates that four-phonon scattering is dominant over three-phonon scattering across the measured temperature range, although higher-order mechanisms are not entirely excluded (Supplemental Material Section 5) [29]. In typical materials, four-phonon scattering often becomes important only at high temperatures, but our finding aligns with prior theoretical calculations that the four-phonon scattering rate for optical phonons in BAs is orders of magnitude larger than the three-phonon scattering rate in a broad temperature range of 77 to 300 K [42]. As shown in Figure 3b, linewidth estimated from molecular dynamics ($\Gamma_{34,MD}$) is larger than our experimental results while that from perturbation theory ($\Gamma_{34,PT}$) is smaller. The trend holds true comparing theoretical linewidth with values above room temperature extrapolated from our experimental results. This could be attributed to the classical equipartition phonon occupations in molecular dynamics and missing scattering channels in perturbation theory [13,27]. Therefore, our results provide valuable reference for future theoretical calculations. Furthermore, the ultralow phonon loss in BAs enables its potential application for devices with a high quality factor [43].

The non-zero constant scattering rate ($B = 0.10$ cm$^{-1}$) highlights the contribution of phonon-isotope and phonon-defect scattering, both of which are temperature-independent. Notably, phonon-defect scattering has been demonstrated to significantly impact the thermal and electrical properties of BAs. For defect concentrations around $10^{19}$ cm$^{-3}$ (~0.03%), the thermal conductivity is reduced to approximately half of its maximum value [35]. In contrast, the phonon-isotope effect has been proven to have a comparatively limited influence on thermal conductivity, with isotopically enriched samples exhibiting only a ~10% increase in thermal conductivity compared to naturally abundant samples[44]. To differentiate the two impurity scattering mechanisms for optical phonons in BAs, we measured the phonon linewidth in samples with different crystal quality. Previous studies have established the relation between defect concentrations, measured by Hall conductivity, electron probe microanalysis, and time-of-flight secondary ion mass spectrometry, and electronic Raman scattering (ERS) strength, characterized by an elevated scattering background and a Fano line shape in the Raman peaks (Supplemental Material Section 6) [2,29,35,45]. To suppress the intrinsic phonon anharmonic scattering, we cooled down the samples to 12 K and measured the Raman spectra. Figure 4a presents the spectra from two samples, one showing a high ERS background and a peak with a Fano line shape while the other shows a low ERS background and a symmetric peak. The relative impurity concentration can be represented by the ERS intensity normalized by the maximum intensity of the phonon peak found in a high-quality sample. We extracted phonon linewidth values from 12 different sample locations with varying ERS intensity. Within the spectral resolution, we found no systematic dependence of the optical phonon linewidth on the ERS intensity, *i.e.,* defect concentration (Figure 4b).

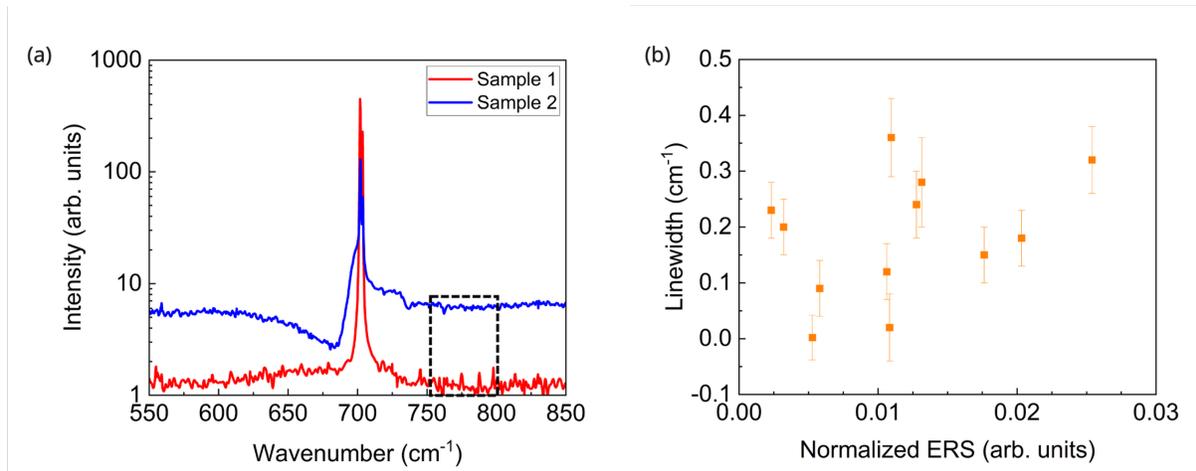

**Figure 4.** Effect of phonon-defect scattering in BAs on phonon linewidth. (a) Representative Raman spectra of BAs samples with different defect concentrations at 12 K. The spectrum for sample 2, which has the higher defect concentration, shows a peak with a Fano line shape. The dashed rectangle highlights the 750-800 cm$^{-1}$ frequency range where the Raman background intensities were integrated to qualitatively estimate defect concentration. (b) The phonon linewidth is found to be independent of the integrated Raman background, i.e. defect concentration. Therefore, within our detection limit, the phonon-defect scattering process does not contribute significantly to the overall decoherence rate of optical phonons in BAs.

Rather, the residue linewidth <0.3 cm$^{-1}$ is consistent with the theoretical expectation from phonon-isotope scattering [27]. Isotopic variation of boron can induce disorder for the optical phonon branch, but has little impact on acoustic phonon branch, whose energy is mainly controlled by arsenic with only one stable isotope $^{75}$As. The disorder relaxes the momentum conservation and enables elastic scattering. The isotope impurity in BAs can reach 20% in naturally abundant samples, and is about 2% in our enriched samples, much larger than the defect concentration in the as synthesized BAs which is around 0.1 – 1% according to ERS intensity [35]. Moreover, the weak electron-phonon coupling strength in BAs could minimize the influence of electronic defects on optical phonons [46,47]. Therefore, we expect that defect-induced linewidth broadening to be smaller than the residue linewidth itself caused by isotope scattering, and thus it is reasonable that phonon-isotope scattering determines the lifetime of optical phonons at low temperatures within the uncertainty of our Raman spectra. Theoretical calculations suggested that the linewidth limited by phonon-isotope scattering is about 0.25 cm$^{-1}$/%, which is much larger than what we observed here. It is possible that the isotope impurities are not uniformly distributed in the crystals, so that the local enrichment determining the phonon frequencies is higher than the nominal value. Currently, the linewidth we observed experimentally is 0.19 cm$^{-1}$ at 100 K, corresponding to a record-high quality factor of 3.7×10$^3$ for optical phonons. A crystal quality with 0.005% defects (1.8×10$^{18}$ cm$^{-3}$, isotope impurity and dopants combined) would support a linewidth around 0.005 cm$^{-1}$ at a reasonably achievable temperature of 20 K, corresponding to a quality factor on the order of 10$^5$ and an ensemble coherence lifetime around 1 ns [27], although such enrichment and growth have not been experimentally demonstrated yet. These values show that properly enriched BAs is a promising semiconductor platform for quantum phononics, and provide motivation for further theoretical and experimental investigation into phonon-isotope scattering.

## IV. CONCLUSIONS

In summary, we systematically studied the scattering mechanisms and observed record-high coherence lifetime for optical phonons in BAs using both IR and Raman spectroscopy. We resolved, for the first time, the small LO-TO splitting and the narrow phonon linewidth inaccessible in previous studies, which enabled detailed investigation of the temperature evolution of phonon properties. We experimentally isolate the four-phonon scattering mechanisms by the quadratic temperature dependence of the optical phonon linewidth, and quantify the intrinsic anharmonicity of optical phonons in BAs in comparison with the first-principles calculations. Furthermore, we found that phonon-defect scattering is negligible impact on optical phonon linewidth in the as-grown high-quality crystals, and observed an exceptionally high quality factor above $3.7 \times 10^3$ for both singly and doubly degenerate modes limited by isotope impurity. These behaviors are in contrast to those of acoustic phonons and thermal transport properties, which are limited by both 3-phonon and higher-order processes and more sensitive to crystalline defects. Our findings offer a clean way to verify calculations involving high-order phonon anharmonicity and encourage experimental efforts in isotope engineering in BAs to achieve ultralong-lived phonon polaritons for mid-infrared photonics. This excellent materials platform may also benefit future studies of polarized phonons with angular momentum and time-reversal symmetry breaking, and time-resolved spectroscopy of finite-momentum phonons whose higher-order scattering processes are also suppressed.

## ACKNOWLEDGMENTS

T.L., W.S., and H.Z. acknowledge the support from Welch Foundation (C-2128). S.S. and H.Z. also acknowledge the Air Force Office of Scientific Research (FA9550-24-1-0135). R.H. acknowledges support by DOE Office of Science Grant No. DE-SC0020334 subaward S6535A. G.Y. was supported by the NSF Grant No. DMR-2104036. C.N. is supported by the NSF Grant No. DMR-2300640. Z.R. acknowledges the support from NSF Grants No. DMR-2425439 and Qorvo, Inc.

**Data and materials availability:** The data that support the findings of this article are openly available [48].

**Author contributions:** H.Z. and Z.R. devised the research. T.L. and W.S. performed infrared spectroscopy. F.P. and A.N. grew the BAs crystals. G.Y., S.S., C.N., S.B. and R.H. performed Raman spectroscopy. T.L., S.S., and H.Z. analyzed the data. All authors contributed to the manuscript writing.